# MODELING AND ANALYZING THE VOCAL TRACT UNDER NORMAL AND STRESSFUL TALKING CONDITIONS

Ismail Shahin[1] and Nazeih Botros[2]
[1]Electrical/Electronics and Computer Engineering Department
University of Sharjah, P. O. Box 27272, Sharjah, United Arab Emirates
[2]Department of Electrical and Computer Engineering
Southern Illinois University at Carbondale, Carbondale, IL 62901-6603, U.S.A.
[1]E-mail: ismail@sharjah.ac.ae
[2]E-mail: botrosn@siu.edu

## ABSTRACT

In this research, we model and analyze the vocal tract under normal and stressful talking conditions. This research answers the question of the degradation in the recognition performance of text-dependent speaker identification under stressful talking conditions. This research can be used (for future research) to improve the recognition performance under stressful talking conditions.

## I. INTRODUCTION: HUMAN SPEECH PRODUCTION MECHANISM

The process of generating speech begins in the lungs. During excitation, muscle contraction forces air out of the lungs through the vocal cords. When the vocal cords remain open, the speech produced is said to be unvoiced and the initial speech spectrum may be modeled as a white noise. On the other hand, when the vocal cords are closed during exhalation, they begin to vibrate, providing an excitation in the form of a periodic train of pulses, the speech produced is said to be voiced speech [1, 2].

The spectrum of either of these excitations is modified by the acoustic cavities formed by the vocal tract. The vocal tract begins at the vocal cords and ends at the lips. The shape of the vocal tract changes continuously which causes the speech sound to be continuously time varying [1, 2]. References [1, 2] have more details about human speech production mechanism.

The conventional division of speech sounds is into consonants and vowels. In a vowel sound, the air in the vocal tract vibrates at frequencies simultaneously. These frequencies are called formant frequencies of the vocal tract. These formant frequencies and their corresponding bandwidths are functions of the shape of the vocal tract [3].

## II. VOCAL TRACT MODEL UNDER NORMAL TALKING STYLE

Under the normal talking style (no stress), the vocal tract can be modeled as shown in Figure 1a. This model can be approximated as shown in Figure 1b. The vocal tract is divided into $p$ number of cylindrical sections which is a fairly close approximation to its actual shape. The vocal tract can be represented by an all-pole transfer function given as [1, 2]:

$$H(z) = \frac{K}{1 + \alpha_1 z^{-1} + ... + \alpha_p z^{-p}} \quad (1)$$

where,
K: is a constant gain.
$\alpha_i$: is the $i$th prediction coefficient which can be calculated using the following

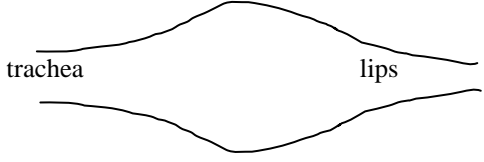

Fig. 1a  Vocal tract under normal talking style

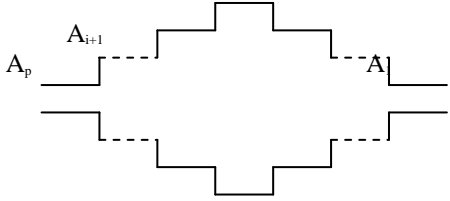

Fig. 1b  Vocal tract approximation under normal talking style

formula if the shape of the vocal tract is known [1]:

$$\alpha_i = \frac{A_i - A_{i+1}}{A_i + A_{i+1}} \quad (2)$$

where,
$A_i$: is the $i$th vocal tract area function.
$A_{i+1}$: is the $(i+1)$th vocal tract area function.

The formant frequencies of the vocal tract and their corresponding bandwidths can be calculated using the following two equations respectively [1]:

$$F_i = \frac{\theta_i f_s}{2\pi} \quad (3)$$

where,

$F_i$ : is the $i$th formant frequency.
$\theta_i$ : is the angle (in radians) of the $i$th pole.
$f_s$ : is the sampling frequency.

$$B_i = \frac{-\ln|z_i| f_s}{\pi} \quad (4)$$

where,

$B_i$: is the bandwidth of the $i$th formant frequency.

$|z_i|$ : is the distance (from the origin) of the $i$th pole.

## III. VOCAL TRACT MODEL UNDER LOUD TALKING STYLE

Under the loud talking style, the vocal tract can be modeled as shown in Figure 2a [4-6]. This model can be approximated as shown in Figure 2b.

Air exits the glottis like a jet and attaches to the nearest wall of the vocal tract. A cavity is formed in the vocal tract because the pressure of the air inside the vocal tract is increased. Vortices of the air are formed as soon as the air passes over the cavity. The bulk of the air continues propagating towards the lips while adhering to the walls of the vocal tract. These vortices produce sound that overlaps with the original sound [4-6].

The $i$th prediction coefficient for the loud talking style can be calculated as:

$$\alpha_i^{Lo} = \frac{A_i^{Lo} - A_{i+1}^{Lo}}{A_i^{Lo} + A_{i+1}^{Lo}} \quad (5)$$

The vocal tract transfer function becomes:

$$H^{Lo}(z) = \frac{K}{1 + \alpha_1^{Lo} z^{-1} + ... + \alpha_p^{Lo} z^{-p}} \quad (6)$$

The locations of the poles of the transfer function are changed to a large extent but the poles are still located inside the unit circle. Therefore, the prediction coefficients under the loud talking style are different to a large extent from those under the normal talking style. Consequently, the cepstral coefficients under the loud talking style are different to a large degree from those under the

normal talking style. Therefore, the cepstral coefficients under the loud

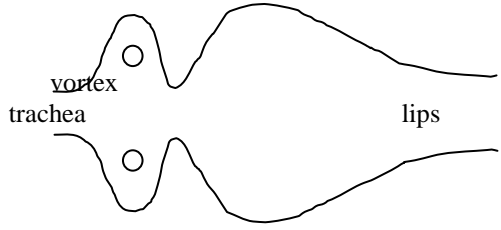

Fig. 2a  Vocal tract under loud talking style

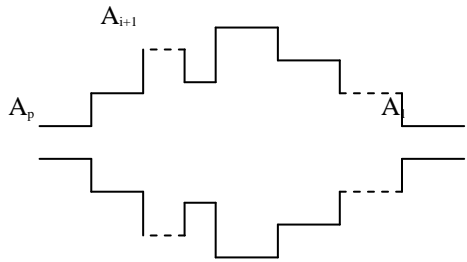

Fig. 2b  Vocal tract approximation under loud talking style

talking style are contaminated with stress components.

Since the formant frequencies of the vocal tract and their corresponding bandwidths are functions of the shape of the vocal tract [3], the formant frequencies and their corresponding bandwidths become:

$$F_i^{Lo} = \frac{\theta_i^{Lo} f_s}{2\pi} \qquad (7)$$

$$B_i^{Lo} = \frac{-\ln\left|z_i^{Lo}\right| f_s}{\pi} \qquad (8)$$

So, the displacement of the formant frequencies of the vocal tract and their corresponding bandwidths under the loud talking style are changed by a large degree.

## IV. VOCAL TRACT MODEL UNDER SHOUT TALKING STYLE

Under the shout talking style, the pressure of the air is increased by a large extent. This increase produces a large cavity which increases the vortices inside the vocal tract. Increasing the vortices yields an increase in the production of sound that overlaps with the original sound [4-6].

The vocal tract transfer function becomes:

$$H^{Sh}(z) = \frac{K}{1 + \alpha_1^{Sh} z^{-1} + ... + \alpha_p^{Sh} z^{-p}} \qquad (9)$$

The locations of the poles of the transfer function are changed to a large extent but the poles are still located inside the unit circle. As in the case of the loud talking style, the prediction coefficients under the shout talking style are different to a large extent from those under the normal talking style. Consequently, the cepstral coefficients under the shout talking style are different to a large degree from those under the normal talking style. Therefore, the cepstral coefficients under the shout talking style are contaminated largely with stress components.

It is known that a part of the sound energy is lost within the vocal tract due to viscous friction, heat conduction, and vibration of the vocal tract wall. This energy loss has significant effects on the vocal tract formant frequencies and their corresponding bandwidths [7].

Since the formant frequencies of the vocal tract and their corresponding bandwidths are functions of the shape of the vocal tract [3], the formant frequencies and their corresponding bandwidths become:

$$F_i^{Sh} = \frac{\theta_i^{Sh} f_s}{2\pi} \quad (10)$$

$$B_i^{Sh} = \frac{-\ln\left|z_i^{Sh}\right| f_s}{\pi} \quad (11)$$

So, the displacement of the formant frequencies of the vocal tract and their corresponding bandwidths under the shout talking style are changed by a large degree.

## V. VOCAL TRACT MODEL UNDER SOFT TALKING STYLE

Under the soft talking style, the pressure of the air is decreased by a small extent. The vocal tract transfer function becomes:

$$H^{So}(z) = \frac{K}{1 + \alpha_1^{So} z^{-1} + \ldots + \alpha_p^{So} z^{-p}} \quad (12)$$

The locations of the poles of the transfer function are changed by a small extent but the poles are still located inside the unit circle. Therefore, the prediction coefficients under the soft talking style are different to a slight range from those under the normal talking style. Consequently, the cepstral coefficients under the soft talking style are different to a small extent from those under the normal talking style. Therefore, the contamination of the cepstral coefficients under the soft talking style is small.

Since the formant frequencies of the vocal tract and their corresponding bandwidths are functions of the shape of the vocal tract [3], the formant frequencies and their corresponding bandwidths become:

$$F_i^{So} = \frac{\theta_i^{So} f_s}{2\pi} \quad (13)$$

$$B_i^{So} = \frac{-\ln\left|z_i^{So}\right| f_s}{\pi} \quad (14)$$

So, the displacement of the formant frequencies of the vocal tract and their corresponding bandwidths under the soft talking style are changed to a small degree.

## VI. VOCAL TRACT MODEL UNDER SLOW TALKING STYLE

Under the slow talking style, the pressure of the air is increased to a small extent. This means that the formation of the vortices inside the vocal tract is small. These small vortices produce a minor sound that overlaps with the original sound [4-6].

The vocal tract transfer function becomes:

$$H^{Sl}(z) = \frac{K}{1 + \alpha_1^{Sl} z^{-1} + \ldots + \alpha_p^{Sl} z^{-p}} \quad (15)$$

The locations of the poles of the transfer function under the slow talking style are close to those under the normal talking style but the poles are still located inside the unit circle. Therefore, the prediction coefficients under the slow talking style are close to those under the normal talking style. Consequently, the cepstral coefficients under the slow talking style are close to those under the normal talking style. Therefore, the contamination of the cepstral coefficients under the slow talking style is minor.

The formant frequencies of the vocal tract and their corresponding bandwidths become:

$$F_i^{Sl} = \frac{\theta_i^{Sl} f_s}{2\pi} \quad (16)$$

$$B_i^{Sl} = \frac{-\ln\left|z_i^{Sl}\right| f_s}{\pi} \quad (17)$$

So, the displacement of the formant frequencies of the vocal tract and their corresponding bandwidths under the slow talking style are close to those under the normal talking style.

## VII. SPEECH DATA BASE

The experiments and tests conducted in this research are performed at Southern Illinois University at Carbondale. Some talking styles are designed to simulate the speech produced by different speakers under real stressful conditions [8, 9]. The talking styles are: normal, shout, slow, loud, and soft. In this research, the data base consists of nine different speakers (three adult males and six adult females) uttering the same word nine times under each talking style.

## VIII. RESULTS

An all-pole transfer function of the vocal tract under any talking style is given as:

$$H^{sty}(z) = \frac{K}{1 + \alpha_1^{sty} z^{-1} + \ldots + \alpha_p^{sty} z^{-p}} \quad (18)$$

The prediction coefficients ($\alpha_1, \alpha_2, \ldots, \alpha_p$) have been calculated using Levinson or Durbin recursion method.

Table I shows the recognition performance under normal and stressful talking conditions using dynamic time warping algorithm [10]. Table II shows the recognition performance under normal and stressful talking conditions using hidden Markov model algorithm [11]. Figures 3 and 4 show the formant frequencies and their corresponding bandwidths for two speakers only.

## IX. DISCUSSION AND CONCLUSIONS

In this research, the following conclusions can be drawn:

1) Comparing the first formant frequencies under the shout, slow, loud, and soft talking styles with the first formant frequencies under the normal talking style, our results show that:

a. The first formant frequencies are displaced to a large degree under the loud talking style. This result is in agreement with the results reported by Wakita and Schulman [7, 12].

b. The first formant frequencies are displaced to a large extent under the shout talking style. This result is in agreement with the results reported by Wakita and Summers [7, 12, 13].

c. The formant frequencies are displaced to a small degree under the soft and slow talking styles.

2) The displacement of the formant frequencies degrades the performance of speaker recognition systems. The higher the displacement, the higher the degradation of recognition performance and vice versa. For example, under the shout talking style, the displacement of the formant frequencies is high which results in high degradation of recognition performance. Another example is that under the slow talking style, the displacement of the formant frequencies is low which results in low degradation of recognition performance.

3) Our results are in agreement with the results reported by Cummings and Clements [14]. Cummings and Clements

reported an extensive investigation of the variations that occur in the glottal excitation of eleven commonly encountered speech styles. Their results showed that the soft and loud talking styles are drastically different from all other styles. Their results also showed that the slow talking style is rarely confused with other styles. Our results are in agreement with their results under the soft and slow talking styles since the recognition performance under these two styles is better to a larger extent in our research. On the other hand, our results are not in agreement with their results under the loud talking style since our results show that the recognition performance under this style is degraded.

4) The highest degradation in the recognition performance happens under the shout talking style. It seems that when speech is contaminated under the shout style, the degree of the contamination is large. This high degree of contamination is caused by the high degree of displacement of the formant frequencies under the shout style.

5) The method of modeling and analyzing the vocal tract under normal and stressful talking conditions that has been used in this research is constrained by the limited amount of data under different talking styles; a comprehensive assessment of the method requires a larger set of test data.

Table I

Recognition rate using dynamic time warping algorithm

| Style | Normal | Shout | Slow | Loud | Soft |
|---|---|---|---|---|---|
| Recognition Rate | 100% | 33% | 51% | 40% | 52% |

Table II

Recognition rate using hidden Markov model algorithm

| Style | Normal | Shout | Slow | Loud | Soft |
|---|---|---|---|---|---|
| Recognition Rate | 90% | 19% | 62% | 38% | 30% |

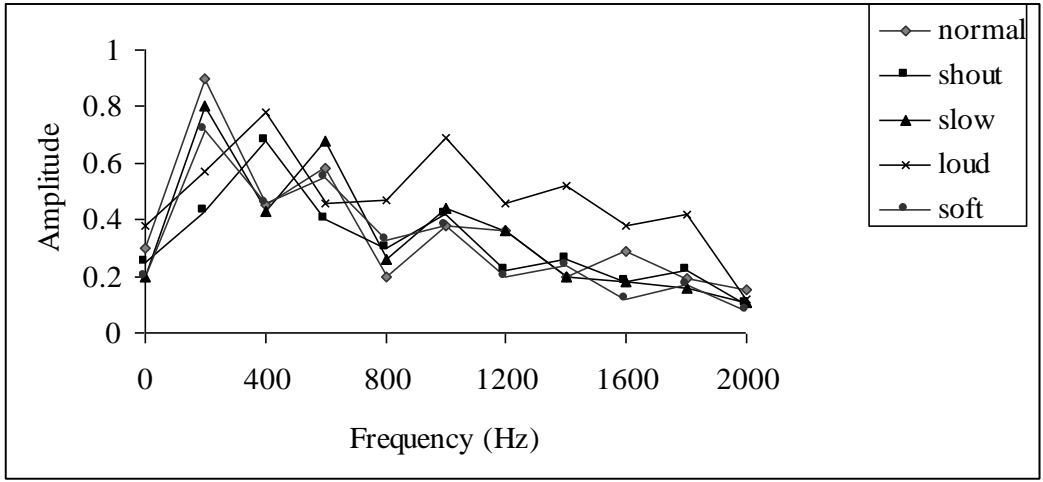

Fig. 3  Formant frequencies of speaker 1

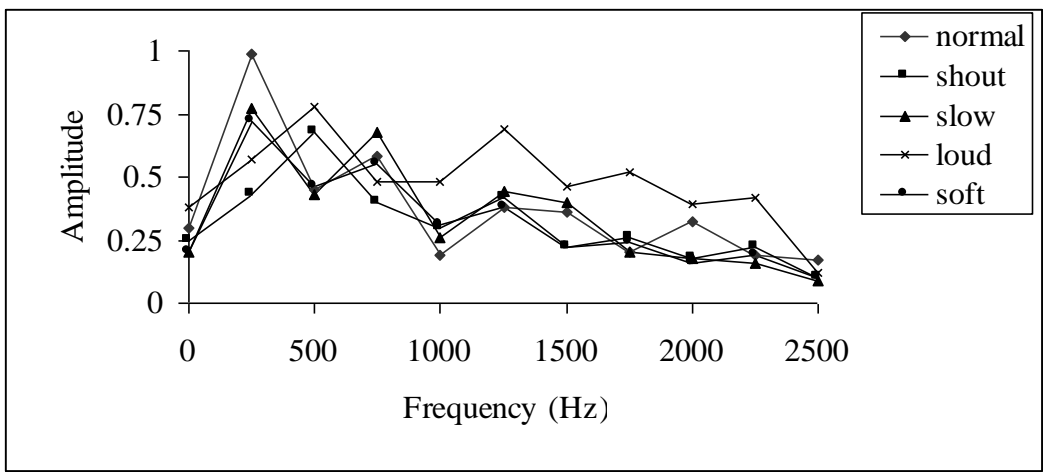

Fig. 4  Formant frequencies of speaker 2